\documentclass[10pt,twocolumn]{article}
\usepackage{graphicx}
\usepackage{epsfig}
\usepackage{amssymb}
\newfont{\normcal}{cmfi10} \newfont{\smallcal}{cmfi10 scaled 900}
\newcommand{\logoLPC}{ {\smallcal\begin{tabular}{r}
 \hline\hline\noalign{\smallskip}
 {\Huge\bf{L}}\hskip-.05em{aboratoire}\hskip-5.7em\raise1.75ex\hbox{
 {\Huge\bf{P}}\hskip-.04em\raise.5ex\hbox{hysique}}\hskip-5.06em\raise-1.25ex\hbox{
 {\Huge\bf{C}}\hskip-.25em\raise-.5ex\hbox{orpusculaire}}\hskip-2.7em\raise4.1ex\hbox{CAEN} 
 \\ \noalign{\smallskip}\hline\hline \end{tabular} }} 
 
\newcommand{\head}[2]{{#1}\hfill{#2}\hfill\raisebox{-4.2ex}{\logoLPC}\bigskip}
\renewcommand{\title}[1]{{\LARGE\bf#1}\bigskip\bigskip}
\def\ifdef#1#2#3{\expandafter\ifx\csname#1\endcsname\relax#3\else#2\fi}
\newcounter{Nad}
\renewcommand{\author}[2]{\ifdef{#2}{}{\expandafter\def\csname#2\endcsname{}
 \newcounter{#2}\stepcounter{Nad}\setcounter{#2}{\theNad}}{#1}$^{\arabic{#2}}$}
\newcommand{\address}[2]{\expandafter\gdef\csname add\roman{#1}\endcsname
 {{\small\it$^{\arabic{#1}}$~#2}\\} }
\newcounter{NAD}\newcommand{\Address}{\stepcounter{NAD}\ifdef{add\roman{NAD}}
 {\csname add\roman{NAD}\endcsname\Address} {\bigskip\bigskip} }
\newcounter{Nth}\renewcommand{\thefootnote}{\alph{footnote}}
\renewcommand{\thanks}[1]
 {\stepcounter{Nth}\expandafter\gdef\csname thanks\roman{Nth}\endcsname
 {\stepcounter{Nth}\footnotetext[\value{NTH}]{~#1}}$^{,}$\footnotemark}
\newcounter{NTH}\newcommand{\Thanks}{\stepcounter{NTH}\ifdef{thanks\roman{NTH}}
  {\csname thanks\roman{NTH}\endcsname\Thanks}
  {\setcounter{footnote}{0}\renewcommand{\thefootnote}{\fnsymbol{footnote}}} }
\renewenvironment{abstract} {\begin{minipage}{15cm}\small{\bf{Abstract:}}}
  {\end{minipage}\bigskip\bigskip}
\newcommand{\PACS}[1]{\vskip1mm {\small{\bf{PACS:}}~#1}}

\let\Ocap\caption \renewcommand{\caption}[1]{\Ocap{\small#1}}

\newenvironment{references}{\begin{thebibliography}{99}}{\end{thebibliography}}
\newcommand{\etal}{{\em et al.}}
\newcommand{\X}[2]{{^{#1}{\mbox{#2}}}}

\newcommand{\PRL}[3]{Phys.\ Rev.\ Lett.\ {\bf#1}, #3 (#2)}
\newcommand{\PRC}[3]{Phys.\ Rev.\ C {\bf#1}, #3 (#2)}

\newcommand{\NIMA}[3]{Nucl.\ Instr.\ Meth.\ A {\bf#1}, #3 (#2)}

\begin{document}
 
\twocolumn[\head{\large\sf (Brief Report, PRC)}{}\begin{center}
 
\title{On the possible detection \\
       of $\X{4}{n}$ events in the breakup of $\X{14}{Be}$}

\author{F.M.~Marqu\'es}	{LPC}\thanks{Electronic address: Marques@lpccaen.in2p3.fr},
\author{N.A.~Orr}	{LPC},
\author{H.~Al~Falou}	{LPC},
\author{G.~Normand}	{LPC},
\author{N.M.~Clarke}	{Bham}
 
\bigskip
 
\address{LPC}{Laboratoire de Physique Corpusculaire,
 IN2P3-CNRS, ENSICAEN et Universit\'e de Caen, F-14050 Caen cedex, France}
 
\address{Bham}{School of Physics and Astronomy,
 University of Birmingham, Birmingham B15~2TT, United Kingdom}
 
\Address

\begin{abstract}
In a recent paper ---F.M.~Marqu\'es \etal, \PRC{65}{2002}{044006}--- 
a new approach to the production and detection of free neutron clusters was 
proposed and applied to data acquired for the breakup of $\X{14}{Be}$. Six 
events that exhibited characteristics consistent with a bound tetraneutron
($\X{4}{n}$) were observed in coincidence with $\X{10}{Be}$ fragments. Here, 
two issues that were not considered in the original paper are addressed: namely 
the signal expected from a low-energy $\X{4}{n}$ resonance, and the detection 
of a bound $\X{4}{n}$ through proccesses other than elastic scattering by a 
proton. Searches complementary to the original study are also briefly noted.
\PACS{21.45.+v; 25.10.+s; 21.10.Gv}
\end{abstract}
 
\end{center}]
 
\Thanks
 
The research, both experimental and theoretical, on free neutron clusters
gained renewed interest following our report of events exhibiting 
characteristics consistent with the detection of
a bound $4n$ cluster liberated in the
breakup of $\X{14}{Be}$ \cite{FMM02}. The approach employed was 
based on the breakup of energetic beams of very neutron-rich nuclei and the 
subsequent detection of the multineutron cluster in liquid scintillator 
modules. The identification of such events was made 
through a comparison of the energy deposited in the 
modules ($E_p$), as generated by the interaction of the putative neutron 
cluster with the protons in the scintillator, with the energy derived from 
the flight time from the breakup target ($E_n$). Multineutron cluster events 
would then be associated with $E_p>E_n$. 
 
As described in Ref.~\cite{FMM02}, the method was applied to data acquired for 
the breakup of intermediate energy (30--50 MeV/nucleon) beams of $\X{11}{Li}$,
$\X{14}{Be}$ and $\X{15}{B}$. In the case of the $\X{14}{Be}$ beam, some 6
events were found exhibiting the characteristics consistent with the production
and detection of a multineutron cluster in coincidence with a
$\X{10}{Be}$ fragment. Much effort was made to estimate 
the effects of pileup; that is the detection for a breakup event of more than 
one neutron in the same module. Three independent approaches were used to 
estimate the rate at which such pileup occurred. It was found that pileup 
could account for at most some 10\% of the observed signal. It was thus 
suggested that at a level of some 2$\sigma$ a signal consistent with
a bound $\X{4}{n}$ liberated in coincidence with $\X{10}{Be}$ had been 
observed.

Not suprisingly these observations have solicited considerable interest.
In particular, a number of theoretical studies were undertaken to investigate 
the conditions under which a bound $4n$ system is permissable (Ref.~\cite{Pie03} 
and references therein). These studies all concluded that given our present 
unerstanding of the $n$-$n$ interaction and the physics associated with
few-body systems (specifically the influence of three-body forces), it is 
impossible to generate a bound $4n$ system. Interestingly, however, the 
calculations of Pieper \cite{Pie03} suggested that it may be possible for 
the tetraneutron to exist as a relatively low-energy resonance.

Some interest has also focussed on the detection process employed in the
experiment. In particular, Bertulani and Sherrill \cite{She03} have 
explored elastic $(\X{4}{n},p)$ scattering, the process proposed in our 
original analysis to be the 
dominant one in the production of the observed events \cite{FMM02}. Using what 
they considered to be reasonable parameters 
based on the very weak binding of the putative $\X{4}{n}$, Bertulani
and Sherrill concluded that the cross-section for backward angle scattering
would be far too low to result in recoil protons with energies in excess of 
that of a single neutron (i.e., $E_p>E_n$).  

In the present paper the detection process and the possibility of a resonant 
$\X{4}{n}$ are addressed. It is argued that the events observed 
in the breakup of $\X{14}{Be}$ may be compatible with a low-energy $\X{4}{n}$ 
resonance because of the high probability of two or more neutrons being
detected in the same detector module. In the case of a bound $\X{4}{n}$,
processes other than elastic proton scattering will contribute to
the detection of events with $E_p>E_n$.


As outlined above, a crucial step in understanding the significance of the
events observed in the breakup of $\X{14}{Be}$ with excess energies deposited 
in the scintillator modules (owing to the finite experimental
resolutions $E_p/E_n>1.4$) was the estimate of the probability of pileup. 
In our original report, Monte-Carlo simulations constituted one of the 
approaches employed \cite{FMM02}. The parameters describing the breakup in the 
simulations were, as described in Ref.~\cite{FMM02}, adjusted so as to 
reproduce for each beam and reaction channel the measured energy, angular and 
multiplicity distributions of the neutrons, and included contributions from 
both the projectile and target. The pileup probabilities so obtained leading 
to events with $E_p/E_n>1.4$ were in line with the observed rates in the case 
of the $\X{11}{Li}$ and $\X{15}{B}$ beam data. In the case of the 
$(\X{14}{Be},\X{10}{Be})$ channel, a probability for pileup to occur of 
about $5\times10^{-4}$ was estimated, whereas the 6 events observed 
corresponded to a probability of $10^{-2}$, a factor $\sim$20 higher.
 
Importantly, these simulations did not include the possibility of any 
correlations occuring between the neutrons. In the light of the suggestion by 
Pieper \cite{Pie03} that the $\X{4}{n}$ system might exist as a low-energy 
resonance, we have reappraised our original estimates to encompass this 
possibility.
 
If the $\X{4}{n}$ system exists as a low-lying resonance\footnote{The reference 
in our original paper of a resonance referred only to a metastable state of
the $\X{4}{n}$ with a halflife long enough to reach the detector array 
($\sim$100~ns).}, the decay in flight will lead to four neutrons with very 
low relative momentum and, consequently, to an increased probability that two 
or more will be detected in the same module. The simulations have, therefore, 
been modified in order to include the decay of a $\X{4}{n}$ resonance. A 
complete treatment of the decay of such a resonance would require an 
examination of all possible decay modes: for example, decay via two dineutrons, 
via a dineutron and two neutrons, etc. Given our lack of knowledge regarding 
the ground-state structure of the $\X{4}{n}$ system, such a comprehensive study 
seems unwarranted.

\begin{figure}[t]
\begin{center}
 \mbox{\psfig{file=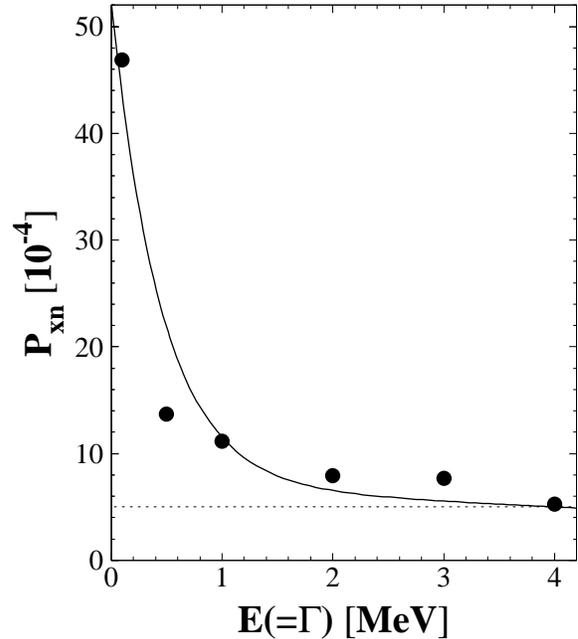,width=8cm}}
 \caption{Pileup probability for different $\X{4}{n}$ resonance energies 
 obtained from the Monte-Carlo simulations described in the text. Here the 
 width of the resonance has been set equal to the resonance energy (see text). 
 The solid line is shown only as a guide to the eye. The dashed line 
 corresponds to the value obtained from the simulations presented in
 Ref.~\protect\cite{FMM02}.} \label{f:ErG}
\end{center}
\end{figure}

In the present context, however, whereby we wish to establish whether the 
signal observed might result from a resonant $\X{4}{n}$, an estimate based on 
four-body phase-space decay is all that is required, as the absence of any 
final-state interactions between the neutrons will lead to the lowest rate 
of pileup of any of the decay scenarios.

Simulations have therefore been carried out for a range of values of the 
energy and width $(E,\Gamma)$ of the resonance, which was parameterised by a 
Breit-Wigner lineshape. As in the original work, both the geometric layout and 
intrinsic detection efficency of the neutron array were taken into account. 
The simulations display for a wide range of energies and widths the expected 
increase of the pileup probability at low resonance energies. For a given 
resonance energy, the results do not depend very strongly on the width. 
We therefore display in Fig.~\ref{f:ErG} the results obtained when the 
width of the resonance is set equal to the resonance energy. We also note that 
the probability for pileup is not very sensitive to the form of the resonance
used to describe the $\X{4}{n}$. As is clearly evident in Fig.~\ref{f:ErG},  
a significant increase of the pileup probability occurs for $E<2$~MeV.

The results presented here should be considered as qualitative. In particular, 
the effects of a resonance will be reduced by the corresponding 
spectroscopic factor for the formation of the $\X{4}{n}$ in the breakup of 
$\X{14}{Be}$. On the other hand, as outline above, the pileup associated with
the decay of a resonant $\X{4}{n}$ would be enhanced by final-state 
interactions between the neutrons. We conclude, therefore, that the events 
reported in our original work could be consistent with the existence of the 
$\X{4}{n}$ system as a resonance, with an energy of the order of 2~MeV or less 
above threshold. We note that similar results are obtained even if the 
resonance is very broad, as suggested in Ref.~\cite{Pie03}.

 
The technique proposed in our original paper \cite{FMM02} for the detection 
of a multineutron cluster liberated in breakup reactions was based on the 
characteristics of $(n,p)$ scattering, the predominant mechanism for the 
detection of neutrons in a liquid scintillator. More specifically, the energy 
recorded in a detector, and attributed to the recoil of a proton ($E_p$), 
cannot exceed the energy of the incident neutron as measured by the 
time-of-flight ($E_n$). Excluding the complications of pileup, events with 
$E_p>E_n$ will, in principle, be generated by the scattering of a heavier 
neutral particle. As described in Ref.~\cite{FMM02}, charged particles were 
vetoed out by the zero-degree Si-CsI telescope and lead shields 
on the entrance windows of the detector modules.

The calculations undertaken by Bertulani and Sherrill \cite{She03} suggest that 
the cross-section for elastic scattering of a very weakly-bound system, such 
as a $\X{4}{n}$, on a proton would be very forward peaked. The elastic 
scattering towards backward angles, responsible for high-energy proton recoils, 
was estimated to be around five orders of magnitude below that of the forward 
angle scattering. This lead to an integrated $(\X{4}{n},p)$ cross-section at 
backward angles of only a few $\mu$b \cite{She03}.

In the context of this result, it should be pointed out that another process 
involving reactions in the scintillator will occur; namely breakup. Indeed, 
the loss of yield in the elastic scattering channel may be attributed 
to breakup. In terms of reactions on the protons in the scintillator, it 
will occur via inelastic scattering or knockout of one of the neutrons. 
Breakup will, of course, also occur on the carbon component of the organic 
scintillator\footnote{As well as on the materials making up the entrance 
window and lead shields of the detector modules.} via absorption or 
diffractive dissociation.

In all cases, the breakup of the $\X{4}{n}$ liberates 4 free neutrons. 
In the energy range considered in the original work, some 10--20~MeV, the 
intrinsic detection efficiency for a neutron is $P_n\sim40\%$. The probability 
for two or more neutrons to be detected, that is to scatter on protons in the 
same detector and deposit energies sufficent for them to be observed, is
$$ 1-\bar{P}_n^4-4P_n\bar{P}_n^3 \ \sim \ 52\% $$
leading, therefore, to a fraction of events with $E_p/E_n>1.4$ similar to that 
of the isotropic elastic scattering scenario sketched in Fig.~3 of 
Ref.~\cite{FMM02}.

Based on the cross-section for breakup of similarly weakly bound systems, such 
as $\X{11}{Li}$, one may expect that for a bound $\X{4}{n}$ to be of the order 
of 1~b or larger. The probability to detect a tetraneutron via breakup in a
DEMON module will, therefore, be of the order of $\sim$0.5~b (i.e., 
$0.52\times1$~b) or more. Given that for energies in the range 10--20~MeV the 
cross-section for $(n,p)$ is of the order of 0.5~b \cite{Wan97}, the 
probability for detecting a bound tetraneutron via breakup is comparable to 
that for detection via $(\X{4}{n},p)$ scattering under the original assumption 
of isotropic scattering.


In conclusion, it has been shown that the events reported in our 
original paper \cite{FMM02} are consistent with the detection of
the $\X{4}{n}$ system as a weakly bound cluster or a low-energy resonance. 
In the case of a bound $\X{4}{n}$, breakup in the
liquid scintillator modules has been identified as the principal detection
process. As outlined above this mechanism is expected to lead to a detection 
probability of the same order or larger than that estimated in 
Ref.~\cite{FMM02} under the assumption of $\X{4}{n}$-$p$ elastic scattering.

In the resonant $\X{4}{n}$ scenario, it has been shown how the in-flight
decay of a low-energy resonance into 4 neutrons with low relative 
momentum increases the probability of two or more of them to be detected in 
the same module, leading to events with $E_p/E_n>1.4$. A resonance with an 
energy of around 2~MeV or less above threshold was shown to be compatible 
with the events observed in Ref.~\cite{FMM02}.

In order to discriminate between a bound and resonant $\X{4}{n}$ we have 
undertaken a breakup measurement with an intense $\X{8}{He}$ beam \cite{ENAM04}.
As outlined in \cite{FMM02}, provided that sufficent statistics are acquired,
the centre-or-mass $\alpha$--$\X{4}{n}$ angular distribution should
provide a means to discriminate between the bound and unbound scenarios.
A complementary technique, which should in principle also be capable of 
distinguishing between a bound and a resonant $\X{4}{n}$, is that of 
missing-mass-type experiments. In the light of presently available beams, 
the $\alpha$ transfer reaction $d(\X{8}{He},\X{6}{Li})4n$ proposed by 
Beaumel \etal\  holds considerable promise \cite{Bea04}. In a similar 
fashion, inelastic scattering, such as the $p(\X{8}{He},p'\alpha)4n$ reaction, 
or at higher energies $\alpha$-knockout, may be employed. Unfortunately,
similar studies are very difficult, if not impossible, in the case of 
$\X{14}{Be}$. If, therefore, the spectroscopic factor for the presence of a 
$\X{4}{n}$ system is very small in $\X{8}{He}$ and significant in $\X{14}{Be}$,
the technique described here and in Ref.~\cite{FMM02} may provide the only
means to access the tetraneutron.



\begin{references}
\bibitem{FMM02} F.M.~Marqu\'es \etal, \PRC{65}{2002}{044006}.
\bibitem{Pie03} S.C.~Pieper, \PRL{90}{2003}{252501}.
\bibitem{She03} B.M.~Sherrill, C.A.~Bertulani, \PRC{69}{2004}{027601}.
\bibitem{Wan97} J.~Wang \etal, \NIMA{397}{1997}{380}.
\bibitem{ENAM04} F.M.~Marqu\'es, Proc.\ of ENAM04, in press. 
\bibitem{Bea04} D.~Beaumel \etal\ (MUST collaboration), private communication.
\end{references}
\end{document}